\documentclass[prd,twocolumn,superscriptaddress,floatfix,amsmath,amssymb]{revtex4-1}

\usepackage[utf8]{inputenc}
\usepackage{amsmath}
\usepackage{amssymb}
\usepackage{physics}
\usepackage{ifthen}
\usepackage{xcolor}
\usepackage{mathtools}
\usepackage{graphicx}
\usepackage{natbib}
\usepackage{bm}
\usepackage{enumerate}
\usepackage{hyperref}
\usepackage{savesym}
\usepackage{makecell}
\usepackage{scalerel}

\renewcommand{\ketbra}[2]{|#1\rangle\!\langle #2|} 
\renewcommand{\k}[1]{\bm{k}_#1}
\newcommand{\Ak}[1]{\annihilation_{\k{#1}}}
\newcommand{\Ck}[1]{\creation_{\k{#1}}}

\newcommand{\ii}{\mathrm{i}}
\newcommand{\expvalstate}[1]{\expval{#1}_{\state}}
\newcommand{\expvalbig}[1]{\big\langle #1 \big\rangle_{\state}}

\newcommand{\commsmall}[2]{[#1,#2]}
\DeclareMathOperator{\Trr}{Tr}

\newcommand{\initialTime}{0} 
\newcommand{\finalTime}{\tau}
 
\newcommand{\dvar}{\mu} 

\newcommand{\HOeigenvalue}{\epsilon_i}
\newcommand{\HTeigenvalue}{\epsilon'_j}

\newcommand{\HObasis}{\{\ket{\epsilon_i}\}}
\newcommand{\HTbasis}{\{|{\epsilon'_j}\rangle\}}

\newcommand{\state}{\hat\rho}
\newcommand{\KMSstate}{\state_{_\beta}}
\newcommand{\unitary}{\hat{U}}
\newcommand{\U}{\hat{U}}
\newcommand{\unitaryDagger}{\unitary^\dagger}
\newcommand{\Hamiltonian}{\hat H}
\newcommand{\HO}{\hat H_\initialTime}
\newcommand{\HT}{\hat H_\finalTime}
\newcommand{\HHT}{\unitaryDagger \HT \unitary} 
\newcommand{\field}{\hat \phi}
\newcommand{\fieldtx}{\field(t,\bm{x})}
\newcommand{\displacement}{\hat D}
\newcommand{\operatorOfWork}{\Delta \hat{\mathcal{U}}}

\newcommand{\annihilation}{\hat{a}}
\newcommand{\creation}{\annihilation^\dagger}

\newcommand{\dk}{\dd[n]{\bm{k}}}

\begin{document}

\title{The First Law of Quantum Field Thermodynamics}
\author{Adam Teixid\'o-Bonfill}
\affiliation{Department of Applied Mathematics, University of Waterloo, Waterloo, Ontario, N2L 3G1, Canada}

\author{Alvaro Ortega}
\affiliation{NonLocal SL, Paseo de la Castellana 259C, Torre de Cristal, Madrid, 28046, Spain}

\author{Eduardo Mart\'in-Mart\'inez}
\affiliation{Institute for Quantum Computing, University of Waterloo, Waterloo, Ontario, N2L 3G1, Canada}
\affiliation{Department of Applied Mathematics, University of Waterloo, Waterloo, Ontario, N2L 3G1, Canada}
\affiliation{Perimeter Institute for Theoretical Physics, 31 Caroline St N, Waterloo, Ontario, N2L 2Y5, Canada}

\begin{abstract}
We study the notion of work fluctuations in quantum field theory, highlighting that the most common definitions used in finite-dimensional quantum systems cannot be applied to quantum field theory (QFT). Then we propose work distributions that are compatible with QFT and we show that they satisfy the first law of thermodynamics up to second moments. We also show how these distributions satisfy Crooks theorem and provide a fully non-perturbative thermodynamic analysis of spacetime localized unitary processes on a quantum field.
\end{abstract}

\maketitle

\section{\textbf{Introduction}} Understanding the role of fluctuations is necessary to formulate the most basic principles of thermodynamics in the context of quantum mechanics~\cite{campisicol, esposito,McKay_2018}. One of the most important and best studied quantities in this context is work of out-of-equilibrium processes~\cite{Baumer2018proposals,referee, tasaki2000jarzynski,Roncaglia_2014,Chiara_2015,Solinas2015fullwork,Solinas2016fullinterference,aharanov1988how,Allahverdyan2014weakwork,Matteo2018quantum,Miller_2017,Sampaio2018Bohmian,Sagawa2012POVM}. However, understanding the notion of work in quantum systems is a notoriously difficult task, since it cannot be associated with an observable~\cite{Talkner2007fluctuationtheorems}. 

One of the most established notions of work fluctuations is the Two-Point Measurement (TPM) work distribution~\cite{Baumer2018proposals,tasaki2000jarzynski, referee}. The TPM work scheme has been used in early pioneering works to explore work distributions for fields confined in cavities~\cite{BRUSCHI2020126601} and free-space~\cite{bartolotta}. However, the TPM distribution is fundamentally incompatible with quantum field theory (QFT)~\cite{Ortega2019work}. One reason is that it relies on projective measurements (PVMs), which are ill-defined in QFT: they break relativistic covariance, introduce causality violations (even in cavity setups) and UV divergences, and have spacetime localization problems, among other issues~\cite{Redhead1995, sorkin, Dowker, Dowker2, borsten2019impossible}. Hence, any definition of work distribution in QFT should not rely on projective measurements in most setups. Since projective measurements are not compatible with QFT, the only way to extract information about the field is by coupling local probes and then measuring those probes~\cite{Unruh,UnruhWald,Fewster, Ortega2019work, bostelmann2020impossible}. 

Furthermore, there is no notion of Gibbs thermality for a quantum field: QFT entropies at constant energy can be divergent and partition functions are not well-defined in QFT in free-space. The proper notion of thermality is captured by the much more general Kubo-Martin-Schwinger (KMS) conditions~\cite{Kubo, schwinger}. Not being able to assume Gibbs thermality makes it harder to prove general theorems about thermal states such as the fluctuation relations \cite{campisicol, esposito,referee,tasaki2000jarzynski}. 

In \cite{Ortega2019work} a PVM-free definition of work distribution for quantum fields (inspired by interferometric experiments) was introduced, showing how it is possible to formulate fluctuation theorems in QFT. Here, we will propose a series of requirements for a work distribution in the context of QFT and identify several different QFT-compatible work distributions which (as we will show) not only satisfy Crooks theorem but also fulfill the first law of thermodynamics on average and variance. Our objective in this manuscript is to find well-defined notions of internal energy difference and work in QFT. We will then apply these notions to find concrete non-perturbative analytic expressions for internal energy difference and work distributions in QFT for arbitrary unitary processes on a free scalar field as a function of their spacetime localization, thus providing closed-form computational tools to apply quantum thermodynamics to QFT. We will work with natural units $\hbar=c=1$.

\section{\textbf{Internal energy in QFT}} \label{sec:Energy}
Before analyzing work distributions, we need to discuss the notion of internal energy difference in QFT. The definition of internal energy difference ($\Delta \mathcal{U}$) as a probability distribution is a tricky concept in quantum thermodynamics. It seems uncontroversial (see, among others~\cite{Baumer2018proposals, Allahverdyan2014weakwork,Perarnau2017nogo,Miller_2017, Sampaio2018Bohmian}) that the expectation of internal energy difference under a unitary process $\U$ is given by
\begin{equation}
\langle\Delta \mathcal{U}\rangle=\Trr(\U\state\U^\dagger\HT) - \Trr(\state\HO), \label{eq:uncontroversialExpectationDU}
\end{equation}
 where $\state$ is the initial state of the system and $\HO$ and $\HT$ are the system Hamiltonians at the start and at the end of the process. However, there seems to be no consensus on how to treat the higher moments of the difference in internal energy. To build a probability distribution for $\Delta\mathcal{U}$ we note that internal energy is a state function. Hence, $\Delta\mathcal{U}$ should only be a function of the state at the start and end of the process and not the path followed. Perhaps one of the most natural ways to build a distribution for internal energy difference is through the following procedure:
\begin{enumerate}
    \item Measure $\HO$ on $\state$, with result $\epsilon_0$ and probability $p_0$.
    \item Take a fresh $\state$, evolve it under $\U$ and  measure $\HT$, with result $\epsilon_\tau$ with probability $p_\tau$.
    \item Then $\Delta \mathcal{U} = \epsilon_\tau - \epsilon_0$ with probability $p_0p_\tau$.
\end{enumerate}
We need to consider a fresh state in step 2 because otherwise $\Delta \mathcal{U}$ would not be path independent (and in fact would coincide with the TPM work distribution~\cite{Baumer2018proposals,tasaki2000jarzynski, referee}). For this definition, the average and variance of $\Delta \mathcal{U}$ are 
\begin{align}
    &\mu_{_{\Delta \mathcal{U}}} = \expvalbig{\HHT} - \expvalbig{\HO}, \label{eq:DUparallelMoments} \\ \nonumber
    &
    \sigma_{_{\Delta \mathcal{U}}}^2 = \expvalbig{\U^\dagger \HT^2 \U} - \expvalbig{\HHT}^2 + \expvalbig{\HO^2} - \expvalbig{\HO}^2.
\end{align}
Notice that the mean coincides with the uncontroversial notion of internal energy difference~\eqref{eq:uncontroversialExpectationDU}.

While this distribution is well-defined for finite-dimensional quantum systems, it is not so in QFT. We show in Appendix~\ref{apx:InternalEnergyDifferenceDistribution} how this distribution presents divergences in its second (and higher moments). Fortunately, the infinities that this distribution presents can be easily renormalized away (for the case of a free theory) defining a new distribution that a) is path independent, b) its expectation is the uncontroversial Eq.~\eqref{eq:uncontroversialExpectationDU}, c) its variance coincides with the finite part of the divergent $\sigma_{_{\Delta \mathcal{U}}}^2$ in Eq.~\eqref{eq:DUparallelMoments}  and d) as we will prove in this manuscript, it satisfies a first law of thermodynamics in mean and variance with the work distributions that are well-defined in QFT. Namely, we define a self-adjoint operator $\operatorOfWork$ from the difference between the Hamiltonians before and after the process $\hat U$:
\begin{equation}
    \operatorOfWork \coloneqq \HHT - \HO. \label{eq:DUoperator}
\end{equation}
The expectation of this operator on the initial state coincides with~\eqref{eq:uncontroversialExpectationDU}, and its variance on KMS states yields the same value as $\sigma_{_{\Delta \mathcal{U}}}^2$ in~\eqref{eq:DUparallelMoments} after subtracting the divergent parts (as shown in Appendix~\ref{apx:InternalEnergyDifferenceDistribution}). We will consider that the moments of $\Delta \mathcal{U}$ are given by \mbox{$\expvalbig{\operatorOfWork^j} =  \big\langle{(\HHT - \HO)^j}\big\rangle_{\hat\rho}$}. Notice that KMS states cannot be eigenstates of~\eqref{eq:DUoperator} unless the process is trivial. 

\section{\textbf{Requirements for a work distribution}} The notion of work in quantum systems should satisfy a series of requirements in order to be  meaningful and useful in the context of QFT. Some of these requirements are features we would want in any work distribution (e.g., be relatable to a classical definition of work, or be associated to a protocol to measure work) and some others are specific of the relativistic nature of QFT. For unitary processes, general thermodynamic considerations yield the following requirements:
\begin{enumerate}
    \item We would like work distributions to satisfy the work fluctuation theorems, Crooks theorem \cite{Crooks1999entropy} and Jarzynski equality \cite{Jarzynski1997nonequilibrium}. These fluctuation theorems relate work distributions of out-of-equilibrium processes to equilibrium quantities~\cite{Jarzynski2008,tasaki2000jarzynski, PhysRevX.8.011019,Mingo2019decomposable,Holmes2019coherentfluctuation,PhysRevE.92.042113,PhysRevX.6.041017}.
    \item We would like work distributions to satisfy the first law of thermodynamics on average and on variance. In particular this would mean that for unitary (i.e., adiabatic) processes
    \begin{align}
    \expvalbig{\operatorOfWork^j} = \expvalbig{W^j} \, , \quad  j=1,2.
    \label{eq:firstlaw}
    \end{align}
    \item There should be an experimental protocol that can measure the work distribution. Besides the obvious practical implications, work should be accessible from an empirical point of view, as it has to be related with a notion of utilizable energy. \end{enumerate}
Furthermore, if we want to build a work distribution that can be covariantly implemented in the context of a fully relativistic QFT, extra requirements are needed:
\begin{enumerate}
    \setcounter{enumi}{3}
    \item The definition of work distribution must not rely on projective measurements. This in turn means that protocols to experimentally measure work performed on a thermodynamic system must not include idealized measurements on the system. This is because PVMs are ill-defined and incompatible with the relativistic nature of QFT \cite{Redhead1995, sorkin, Dowker, Dowker2,borsten2019impossible,Ortega2019work}. This requirement excludes one of the most common definitions of work distribution, the TPM scheme~\cite{Baumer2018proposals,tasaki2000jarzynski, referee}. Work distributions that do not rely on projective measurements can be built coupling ancillary probes to the system (for QFT, in a covariant way~\cite{Tales1,Tales2}), and then extracting information about the work statistics by projectively measuring the probe. Probe-based work distributions have been introduced in non-relativistic contexts \cite{Roncaglia_2014,Chiara_2015,Solinas2015fullwork,Solinas2016fullinterference, De_Chiara2018ancilla} and have been successfully exported to the covariant formalism of QFT \cite{Ortega2019work}.
    \item Work distributions should be well-defined for processes involving thermal states. This is fundamental to satisfy fluctuation theorems. However, satisfying this requirement is non-trivial in QFT, since Gibbs states are, in general, not well-defined (in general $e^{-\beta \Hamiltonian}$ is not a trace class operator and therefore the partition function is ill-defined). In QFT we need to work with the (more general) notion of KMS  thermality~\cite{Kubo,schwinger}. A work distribution suitable for QFT should therefore be able to handle KMS states even when the notion of Gibbs thermality breaks.
    \end{enumerate}

Remarkably, the no-go theorem \cite{Perarnau2017nogo} seems to indicate that there may be no quantum probability distribution for work that satisfies requirements 1 and 2. Therefore we need to allow for work distributions to be quasiprobabilities (not necessarily positive semi-definite) distributions. Indeed work distribution proposals in the literature that satisfy requirements 1 and 2 are quasiprobabilities  (see e.g., \cite{Allahverdyan2014weakwork,Solinas2015fullwork,Matteo2018quantum}). 

As we will see below, two previously introduced work quasi-probability distributions that are well-known in the literature fulfill these conditions: 1) the Allahverdyan-Terletsky-Margenau-Hill work distribution (ATMH) \cite{Allahverdyan2014weakwork, Baumer2018proposals}, and 2) the full-counting statistics (FCS) work distribution \cite{Solinas2015fullwork,Baumer2018proposals}. A third distribution was recently introduced precisely in the context of QFT \cite{Ortega2019work} through a generalization of the Ramsey scheme protocol devised in \cite{mazzola2013measuring, dorner2013extracting} beyond the TPM distribution. The Ramsey-scheme distribution has the advantage that requirement 3 (the experimental protocol) can be associated to measurements with Unruh-DeWitt detectors \cite{Unruh,DeWitt1979,UnruhWald}, which are good models for measurements of quantum fields (and can be connected with the light-matter interaction~\cite{Mart_n_Mart_nez_2013,Pozas2016,Pablo}) without requiring any PVMs. However, this distribution does not fully satisfy requirement 2, but it does satisfy it for a class of states including those that commute with the field Hamiltonian (of which KMS states are a subset). Interestingly, we will also discuss that the real part of the Ramsey scheme work distribution does satisfy all the requirements (including 2 for all states) and we will show that it in fact coincides exactly with the ATMH distribution. Consequently, while previous proposals to experimentally measure the ATMH work distribution required idealized measurements~\cite{Matteo2018quantum,Baumer2018proposals}, it is possible to measure the ATMH distribution without PVMs on the system.

\section{Examples of Work distributions compatible with QFT}
\subsection{\textbf{Ramsey scheme work distribution}} This is the distribution introduced in QFT in \cite{Ortega2019work}.
In particular, the protocol yielding the Ramsey scheme work distribution for a unitary operation $\unitary$ on a quantum system is as follows~\cite{mazzola2013measuring,dorner2013extracting}: we begin with the initial state of the quantum system $\state$, plus an auxiliary qubit playing the role of a probe,  $\state \otimes \ketbra{0}{0}$. Applying a Hadamard on the qubit results on $\state_0 = \state \otimes \ketbra{+}{+}$. We then apply the controlled unitary evolution
\begin{equation}
    \hat{M}_\mu = \unitary e^{-\ii\mu\HO} \otimes \ketbra{0}{0} + e^{-\ii\mu\HT} \unitary \otimes \ketbra{1}{1}.
\end{equation}
Finally, we apply a second Hadamard to the qubit. At the end of this procedure, the reduced state of the qubit can be written as \mbox{$\state_\mu = \frac{1}{2}[\openone + \Re(\widetilde{P}_{\textsc{rs}}(\mu)) \hat{\sigma}_z + \Im(\widetilde{P}_{\textsc{rs}}(\mu)) \hat{\sigma}_y]$}, where 
\begin{equation}
    \widetilde{P}_{\textsc{rs}}(\dvar) = \expvalstate{\unitaryDagger e^{\ii\dvar\HT} \unitary e^{-\ii\dvar\HO}}.
    \label{eq:RamseyCharacteristicWork}
\end{equation}
The Ramsey scheme distribution defined in \cite{Ortega2019work} can be thought of as a particular case of the more general Kirkwood-Dirac quasi-probability~\cite{Kirkwood1933,Dirac1945,Dressel_2015,YungerHalpern2017,Yunger2018Quasiprob} of a bounded operator $\hat O$ and two orthonormal bases $\{\ket{f}\}$ and $\{\ket{a}\}$, \mbox{$p_{\textsc{kd}}(a,f) \coloneqq \braket{f}{a}\!\!\!\mel{a}{\hat O}{f} = \langle\unitaryDagger \ketbra{\HTeigenvalue}{\HTeigenvalue} \unitary \ketbra{\HOeigenvalue}{\HOeigenvalue}\rangle_{\state}$}, particularizing to the joint distribution of internal energy before and after a process (taking  $\HObasis$ and $\HTbasis$ as the eigenbases of $\HO$ and $\HT$ respectively, and $\hat O = \hat\rho$). We can define a work distribution as
\begin{equation}
P_\textsc{kd}(W)\coloneqq\sum_{i,j} \langle\unitaryDagger \ketbra{\HTeigenvalue}{\HTeigenvalue} \unitary \ketbra{\HOeigenvalue}{\HOeigenvalue}\rangle_{\state}\,\delta(W-(\HTeigenvalue-\HOeigenvalue)).
\label{KD_Dist}
\end{equation} If  we Fourier transform \eqref{KD_Dist} we obtain the characteristic function  $\widetilde{P}_{\textsc{rs}}$ in \eqref{eq:RamseyCharacteristicWork}:
\begin{align}
     \sum_{i,j} e^{\ii\dvar(\HTeigenvalue-\HOeigenvalue)} \expval{\unitaryDagger \ketbra{\HTeigenvalue}{\HTeigenvalue} \unitary \ketbra{\HOeigenvalue}{\HOeigenvalue}}_{\hat \rho}\!\!\! =\! \expval{\unitaryDagger e^{\ii\dvar\HT}  \unitary e^{-\ii\dvar\HO}}_{\hat \rho}\!\!.
\end{align}
$\widetilde{P}_{\textsc{rs}}$ coincides with the characteristic function of the TPM work distribution when the TPM work distribution is well-defined and the initial state commutes with the initial Hamiltonian \cite{Talkner2007fluctuationtheorems}. Since we are going to move beyond those cases for KMS quantum field states, we choose to define the Ramsey scheme work distribution as the inverse Fourier transform of $\widetilde{P}_{\textsc{rs}}$, which always exists even in QFT \cite{Ortega2019work}, that is:
\begin{equation}
    P_\textsc{rs}(W)\coloneqq \mathcal{F}^{-1}\{\tilde P_{\textsc{rs}}\}(W)= \frac{1}{2\pi} \int_{\mathbb{R}}\!\widetilde{P}_{\textsc{rs}}(\mu)  e^{-\ii W\mu} \differential\mu\,. 
    \label{eq:RamseySchemeDistributionDef}
\end{equation}
Note that the Ramsey scheme work distribution is a quasiprobability distribution which can take complex values outside diagonal states in the $\HObasis$ basis. The protocol to measure $P_\textsc{rs}$ has already been experimentally implemented~\cite{Batalhao2014experimental,Cetina96}, and is easily implementable in a plethora of scenarios where probes can be coupled to the system, including QFT using particle detectors~\cite{Ortega2019work}, therefore satisfying requirement 3. 
$P_\textsc{rs}$ was used in \cite{Ortega2019work} as a way to define work in quantum fields, since its measurement protocol does not rely on projective measurements and it can handle KMS states, thus satisfying requirements 4 and 5. 

In particular, \cite{Ortega2019work} shows that (unlike the TPM distribution) $P_\textsc{rs}$ can be computed for a general spacetime localized unitary operation on a quantum scalar field given by
\begin{equation}
    \hat{U}=\mathcal{T} \exp {-\ii \int_{\mathbb{R}} \differential t \int_{\mathbb{R}^n}\!\differential^n \bm{x} \,\hat{O}(t, \boldsymbol{x})}\,,
    \label{eq:general_unitary_QFT}
\end{equation}
where $\mathcal{T}$ is the time-ordering operator, $(t,\bm x)$ is an inertial quantization frame in a  \mbox{$(n+1)$-dimensional} Minkowski spacetime, and $\hat O(t,\bm x)$ is an arbitrary linear combination of the quantum field amplitude $\hat\phi(t,\bm x)$ and its canonical momentum $\hat\pi(t,\bm x)$ (which for a free theory represents any element of the algebra of field observables). The work distribution can be computed in terms of the field's Wightman $n$-point functions~\cite{Ortega2019work}, and it can be easily evaluated for KMS states. In this context, the qubit probe is substituted by an Unruh-DeWitt particle detector~\cite{Unruh,DeWitt1979,UnruhWald}, which are good models for measurements of quantum fields that have experimental realizations in the light-matter interaction~\cite{Mart_n_Mart_nez_2013,Pozas2016,Pablo}.

\subsection{\textbf{ATMH work distribution}} This work distribution was first proposed by Allahverdyan \cite{Allahverdyan2014weakwork}, and is directly related to  the Terletsky-Margenau-Hill quasi-probability~\cite{Terletsky1937,Margenau1961correlation} 
particularized to a joint distribution of internal energy before and after a process \cite{Baumer2018proposals}:
\begin{equation}
    p_{\textsc{tmh}}(i,j) = \Re \expvalstate{\unitaryDagger \ketbra{\HTeigenvalue}{\HTeigenvalue} \unitary \ketbra{\HOeigenvalue}{\HOeigenvalue}}
    \label{eq:Margenau-Hill probability}.
\end{equation}
The ATMH work quasi-probability distribution has been defined outside of QFT as  $P_{\textsc{atmh}}(W) = \sum_{ij}p_{\textsc{tmh}}(i,j)\delta(W-(\HTeigenvalue-\HOeigenvalue))$. In that context, one can Fourier transform it to obtain the characteristic function, $\widetilde{P}_{\textsc{atmh}}(\dvar) = \sum_{i,j} e^{\ii\dvar(\HTeigenvalue-\HOeigenvalue)} \Re \langle\unitaryDagger \ketbra{\HTeigenvalue}{\HTeigenvalue} \unitary \ketbra{\HOeigenvalue}{\HOeigenvalue}\rangle_{\state}$ so that
\begin{align}
    \label{eq:ATMHCharacteristic}
    \widetilde{P}_{\textsc{atmh}}(\dvar) = \frac{1}{2} \expvalstate{\unitaryDagger e^{\ii\dvar\HT}  \unitary e^{-\ii\dvar\HO} + e^{-\ii\dvar\HO} \unitaryDagger e^{\ii\dvar\HT} \unitary}.
\end{align}
The ATMH distribution can be implemented as a weak measurement, with the protocol described in \cite{Matteo2018quantum, Baumer2018proposals}. The protocol applies projective measurements on the system, thus, with this implementation it would not satisfy requirement 4. However, comparing Eqs.~\eqref{eq:RamseyCharacteristicWork} and \eqref{eq:ATMHCharacteristic} we see that \mbox{$\widetilde{P}_{\textsc{atmh}}(\dvar) = \frac{1}{2}(\widetilde{P}_{\textsc{rs}}(\dvar) + [\widetilde{P}_{\textsc{rs}}(-\dvar)]^*)$}. Then, since $\mathcal{F}\{(P_{\textsc{rs}})^*\}(\mu)=[\widetilde{P}_{\textsc{rs}}(-\dvar)]^*$, we conclude that \mbox{$P_{\textsc{atmh}} = \Re{P_{\textsc{rs}}}$}. Note that this equality was expected because the Terletsky-Margenau-Hill distribution is the real part of the Kirkwood-Dirac distribution~\cite{Dressel_2015,YungerHalpern2017,Yunger2018Quasiprob}. Hence, the ATMH distribution satisfies 3 through the detector-based Ramsey scheme protocol that defines $P_{\textsc{rs}}$.

It also satisfies 4 if we define the work distribution as the inverse Fourier transform of the characteristic function~\eqref{eq:ATMHCharacteristic}, so that it is well defined even in QFT: $P_\textsc{atmh}\coloneqq \mathcal{F}^{-1}\{\tilde P_{\textsc{atmh}}\}$. We can verify that requirement 5 is also satisfied comparing \eqref{eq:RamseyCharacteristicWork} and \eqref{eq:ATMHCharacteristic}, noticing that \mbox{$[\state,\HO]=0\Rightarrow P_\textsc{rs}=P_{\textsc{atmh}}$} and noting that KMS states commute with $\HO$.

\subsection{\textbf{Full-counting-statistics work distribution}} This work distribution is derived from joint distributions for non-commuting observables \cite{Levitov1993,Levitov1996,Yu2003,Hofer2017quasiprobability}. Its characteristic function is \cite{Solinas2015fullwork}
\begin{equation} 
    \widetilde{P}_{\textsc{fcs}}(\dvar) =\expvalstate{e^{-\ii\frac{\dvar}{2}\HO} \unitaryDagger e^{\ii\dvar\HT} \unitary e^{-\ii\frac{\dvar}{2}\HO}}. 
\label{eq:FCSCharacteristic}
\end{equation}
There are proposals to measure ${P}_{\textsc{fcs}}$ without projective measurements on the system~\cite{Solinas2016fullinterference}, therefore fulfilling requirements 3 and 4. Comparing \eqref{eq:RamseyCharacteristicWork}, \eqref{eq:ATMHCharacteristic} and \eqref{eq:FCSCharacteristic}, we get \mbox{$[\state,\HO]=0\Rightarrow {P}_{\textsc{fcs}}=P_\textsc{rs}=P_{\textsc{atmh}}$}. Since KMS states commute with $\HO$, then ${P}_{\textsc{fcs}}$ also satisfies requirement 5. Note that for finite-dimensional systems and Gibbs states the coincidence of these distributions with the TPM distribution was already known~\cite{Baumer2018proposals, mazzola2013measuring,dorner2013extracting}. 

\section{\textbf{Fluctuation theorems}} We are going to verify that $P_\textsc{rs},P_{\textsc{atmh}},{P}_{\textsc{fcs}}$ satisfy requirement 1. Let us consider initial thermal states, which in QFT are KMS states. A KMS state \cite{Kubo,schwinger} of (inverse) temperature $\beta$ with respect to a Hamiltonian $\Hamiltonian$ and a time direction $\partial_t$ is a state $\KMSstate$ for which all pairs of bounded operators $\hat A, \hat B$ satisfy:
\begin{enumerate}[i.]
    \item The expectation values $\langle\hat A \hat B(t)\rangle_{\KMSstate}$ and $\langle\hat B(t)\hat A\rangle_{\KMSstate}$ are boundary values of some complex functions $\langle\hat A \hat B(z)\rangle_{\KMSstate}$ and $\langle\hat B(z)\hat A\rangle_{\KMSstate}$ holomorphic in the strips $0 < \Im(z) < \beta$ and $-\beta < \Im(z) < 0$, respectively;
    \item  $\langle\hat B(t)\hat A\rangle_{\KMSstate} =   \langle\hat A \hat B(t+\ii \beta)\rangle_{\KMSstate} $, \label{item:secondKMS}
\end{enumerate}
where $\hat B(t)\coloneqq e^{\ii t\Hamiltonian} \hat B e^{-\ii t\Hamiltonian}$.
Since ${P}_{\textsc{fcs}}=P_\textsc{rs}=P_{\textsc{atmh}}$ for KMS states we can prove simultaneously that the three work distributions satisfy fluctuation theorems. For concreteness, we choose the characteristic function \mbox{$\widetilde{P}_\textsc{w}\equiv \widetilde{P}_\textsc{rs}$} in \eqref{eq:RamseyCharacteristicWork}.

Crooks theorem~\cite{Crooks1999entropy} relates $P_\textsc{w}$ for a unitary process $\hat U$ and an initial KMS state of $\HO$ with the work distribution of the time-reversed process ($P_{\text{rev}}$) implemented by $\unitaryDagger$ on a KMS state of $\HT$. Crooks theorem states~\cite{sai2016thermodynamics,Gong_2015,tasaki2000jarzynski} 
\begin{equation}
    \frac{P_\textsc{w}(W)}{P_{\text{rev}}(- W)} = e^{\beta (W-\Delta F)} \Leftrightarrow  \frac{\widetilde{P}_\textsc{w}(\dvar + \ii \beta)}{\widetilde{P}_{\text{rev}}(-\dvar) } = e^{-\beta \Delta F}. 
    \label{eq:CrooksTheorem}
\end{equation}
Here $\Delta F$ is the change in free energy, and for general KMS states it is defined as \mbox{$\Delta F = -\frac{1}{\beta}\ln{\langle e^{-\beta \HT}e^{\beta \HO}\rangle_{\KMSstate}}$}, where  $\langle e^{-\beta \HT}e^{\beta \HO}\rangle_{\KMSstate}$ is the ratio of partition functions $Z_\tau/Z_0$ that, when Gibbs states are well defined, corresponds to $\Tr e^{-\beta \HT}/\Tr e^{-\beta \HO}$. The proof of Crooks theorem simplifies when $\HT = \HO$. Then $\Delta F = 0$ and we should find $\widetilde{P}_\textsc{w}(\dvar + \ii \beta) = \widetilde{P}_{\text{rev}}(-\dvar)$. We obtain this applying the (ii) KMS condition on $\widetilde{P}_\textsc{w}(\dvar + \ii \beta)$, 
\begin{equation}
\begin{split}
        \widetilde{P}_\textsc{w}(\dvar+\ii \beta) &= \expval{\unitaryDagger e^{\ii (\dvar+ \ii \beta)\HO} \unitary e^{-\ii (\dvar+ \ii \beta)\HO} }_{\KMSstate} \\
        &= \expval{e^{\ii\dvar\HO} \unitary e^{-\ii\dvar\HO} \unitaryDagger }_{\KMSstate} ,
\end{split}
\end{equation}
which equals $\widetilde{P}_{\text{rev}}(-\dvar)$ due to $\big[\KMSstate,\HO\big] = 0$. This proves Crooks theorem for $\HT = \HO$, which in turn implies Jarzynski equality~\cite{Jarzynski1997nonequilibrium,sai2016thermodynamics}, $\langle{e^{-\beta W}}\rangle_{\KMSstate} = e^{-\beta \Delta F}$. 

We can also sketch a proof of Crooks theorem for the most general situation of non-cyclic unitary process $\unitary$ that starts with a Hamiltonian $\HO$ and finishes with $\HT$. 

Concretely, we consider a general unitary non-equilibrium process for which the initial state is a KMS state of $\HO$ for the forward process, and a KMS state of $\HT$ for the reverse process. We denote them as $\KMSstate^\initialTime$ and $\KMSstate^\finalTime$ respectively, with $\beta$ the inverse KMS temperature.
Crooks theorem \eqref{eq:CrooksTheorem} relates the functions
\begin{align}
    &\widetilde{P}_{\textsc{w}}(\dvar + \ii \beta) = \expval{\unitaryDagger e^{\ii(\dvar + \ii \beta)\HT} \unitary e^{-\ii(\dvar + \ii \beta)\HO}}_{\KMSstate^\initialTime},\\
    &\widetilde{P}_{\text{rev}}(-\dvar) = \expval{\unitary e^{-\ii\dvar\HO} \unitaryDagger e^{\ii\dvar\HT}}_{\KMSstate^\finalTime}.\label{eq:characteristic function direct and reverse}
\end{align}
To prove the relation we start at $\widetilde{P}_{\textsc{w}}(\dvar + \ii \beta)$ and using the KMS condition we transform it into $\widetilde{P}_{\text{rev}}(-\dvar)e^{-\beta\Delta F}$. To do this, let us introduce a new variable, $\dvar'$, and a new function
\begin{equation}
    \widetilde{P}'(\dvar, \dvar') =  \expval{\unitaryDagger e^{\ii\dvar\HT} e^{-\ii\dvar\HO} e^{\ii\dvar'\HO} \unitary e^{-\ii\dvar'\HO}}_{\KMSstate^\initialTime},
\end{equation}
such that $\widetilde{P}'(\dvar, \dvar) = \widetilde{P}_{\textsc{w}}(\dvar)$. Let us now apply the second KMS property with $t=\mu'$,
\begin{align}
    &\widetilde{P}'(\dvar+\ii \beta, \dvar' + \ii \beta) \nonumber \\ 
    &= \expval{e^{\ii\dvar'\HO} \unitary e^{-\ii\dvar'\HO} \unitaryDagger e^{\ii(\dvar+ \ii \beta)\HT} e^{-\ii(\dvar+ \ii \beta)\HO} }_{\KMSstate^\initialTime}. 
\end{align}
We recover the characteristic function, by equating $\dvar$ and $\dvar'$,
\begin{equation}
    \widetilde{P}_{\textsc{w}}(\dvar + \ii \beta) = \expval{\unitary e^{-\ii\dvar\HO}\unitaryDagger e^{\ii\dvar\HT} e^{-\beta\HT} e^{\beta\HO}}_{\KMSstate^\initialTime} , \label{eq:forward characteristic function KMS applied}
\end{equation}
 where we have simplified using $[\KMSstate^\initialTime,\HO] = 0$. If we prove that \mbox{$e^{-\beta\HT} e^{\beta\HO} \KMSstate^\initialTime  = \KMSstate^\finalTime \langle e^{-\beta\HT} e^{\beta\HO} \rangle_{\KMSstate^\initialTime}$}, then we get Crooks theorem in the following form
\begin{equation}
    \widetilde{P}_{\textsc{w}}(\dvar + \ii \beta) = \widetilde{P}_{\text{rev}}(-\dvar) \big\langle e^{-\beta\HT} e^{\beta\HO} \big\rangle_{\KMSstate^\initialTime}\,,
\end{equation}
from comparing equations \eqref{eq:characteristic function direct and reverse} and \eqref{eq:forward characteristic function KMS applied}. To get the original Crooks theorem we specify 
\begin{equation}
    e^{-\beta \Delta F} = \big\langle e^{-\beta\HT} e^{\beta\HO} \big\rangle_{\KMSstate^\initialTime},
\end{equation} which recovers \mbox{$e^{-\beta \Delta F} = \Trr{e^{-\beta\HT}}/\Trr{e^{-\beta\HO}}$} for Gibbs states. Notice that the ratio of partition functions is also well defined when Gibbs states are replaced by KMS states (see e.g.,~\cite{Simidzija2018harvesting}).  To complete the proof we need to show that $e^{-\beta\HT} e^{\beta\HO} \KMSstate^\initialTime$ is an unnormalized KMS thermal state of $\HT$. We show it when Gibbs states are well defined,
\begin{align}
    e^{-\beta\HT} e^{\beta\HO} \KMSstate^\initialTime &= e^{-\beta\HT} e^{\beta\HO} \frac{e^{-\beta\HO}}{\Tr e^{-\beta\HO}} \nonumber \\ 
    &= \frac{e^{-\beta\HT}}{\Tr e^{-\beta\HT}} \frac{\Tr e^{-\beta\HT}}{\Tr e^{-\beta\HO}}  \\
    &= \KMSstate^\finalTime e^{-\beta \Delta F}, \nonumber
    \label{eq:unnormalizedKMS}
\end{align}
where taking traces at the start and the end gives $\Trr\{e^{-\beta\HT} e^{\beta\HO} \KMSstate^\initialTime\} = e^{-\beta \Delta F}$. It can be argued, in the same manner as in \cite{Simidzija2018harvesting}, that the proof of Crooks theorem for any unitary process follows for full KMS thermality, but we have left the subtleties of this claim out of this work.

\section{\textbf{The first law of (QFT) thermodynamics}}
\label{sec:FirstLaw}
We now analyze whether the three work distributions satisfy the first law both on average and for the second moments. The moments of work are computed from the respective characteristic functions,
\begin{equation}
\expvalstate{W^j} = \ii^{-j} \eval{\frac{\differential^j}{\differential\dvar^j}\widetilde{P}_\textsc{w}(\dvar)}_{\dvar=0}\,.
\label{eq:moments}
\end{equation}

It is straightforward to generalize for QFT the proofs in~\cite{Allahverdyan2014weakwork,De_Chiara2018ancilla} that $P_{\textsc{atmh}}$ and $P_{\textsc{fcs}}$ satisfy  \mbox{$\langle{W_\textsc{atmh}^j}\rangle_{\state} =\langle{\operatorOfWork^j}\rangle_{\state} = \langle{W_\textsc{fcs}^j}\rangle_{\state}$} for $j=1,2$. Thus these two distributions obey the first law of thermodynamics for the first two moments.

Notice that for KMS states all work distributions considered in this paper coincide exactly, which automatically implies that the Ramsey scheme work distribution would also satisfy the first law on average and variance for thermal states. We will compute now the deviation from equation~\eqref{eq:firstlaw} in second moments for $P_\textsc{rs}$ when considering general states. To do this, we will compute the first two moments of the Ramsey scheme work distribution and compare them with the moments of internal energy difference. 

We compute the moments of the Ramsey scheme work from its characteristic function in Eq. \eqref{eq:RamseyCharacteristicWork}. From \eqref{eq:moments}, we get that 
\begin{align}
    \expvalbig{W_{\textsc{rs}}} &= \expvalstate{\unitaryDagger \HT \unitary - \HO}, \\
    \expvalstate{W^2_{\textsc{rs}}} &= \expvalstate{\unitaryDagger \HT^2 \unitary - 2\unitaryDagger \HT \unitary\HO + \HO^2}.
\end{align}
Recall that from the definition of renormalized internal energy difference operator we get for the first and second moments 
\begin{align}
    \expvalbig{\operatorOfWork} &= \Big\langle{\unitaryDagger\HT\unitary - \HO}\Big\rangle_{\hat{\rho}}\,,
    \label{eq:momentsofDU}\\
    \expvalbig{\operatorOfWork^2} \!&= \expvalstate{\unitaryDagger \HT^2 \unitary - \unitaryDagger \HT \unitary\HO - \HO\unitaryDagger \HT \unitary + \HO^2}\!.\nonumber
\end{align}
Therefore, for the Ramsey scheme work distribution we obtain
\begin{align}
    \expvalbig{\operatorOfWork} &= \expvalstate{W_\textsc{rs}}\,, \\
    \expvalbig{\operatorOfWork^2} &= \expvalstate{W_\textsc{rs}^2} + \expvalbig{[\HHT,\HO]}\, .
\end{align}

In this sense the first law for $P_\textsc{rs}$ in  second moments holds in states satisfying $\expvalbig{[\unitaryDagger\HT\unitary,\HO]} = 0$.  This includes KMS states and all initial states which commute with the initial Hamiltonian (this is shown by expressing the difference as a trace and applying the cyclic property). Also, since $[\HHT,\HO]$ is an anti-self-adjoint operator, its expectation is purely imaginary, which tells us that the real part of the Ramsey distribution  satisfies the first law in first and second moments for all states. This is not surprising since we already showed that the real part of $P_\textsc{rs}$ is exactly equal to  $P_{\textsc{atmh}}$.

For third or higher moments the first law fails for all distributions even for initial thermal states. We show explicitly the magnitude of this violation for the case of a free scalar quantum field in the table of Appendix~\ref{apx:Moments_of_Work_and_EnergyDifference_of_KMS_Table}. This is not unexpected since the internal energy probability distribution does not satisfy the fluctuation theorems (see Appendix \ref{apx:Moments_of_Work_and_EnergyDifference_of_KMS_Table} for details) and therefore it cannot coincide exactly with any work distribution that satisfies Crooks theorem. This means that for a high enough moment the first law cannot be satisfied for thermal states and adiabatic processes.

\section{\textbf{Work in quantum fields}} We have seen that $P_{\textsc{rs}}$, $P_{\textsc{atmh}}$, $P_{\textsc{fcs}}$ satisfy all requirements to be well-defined for QFT. In this section we will compute (non-perturbatively) the characteristic functions for work and internal energy difference when a unitary of the form~\eqref{eq:general_unitary_QFT} is applied on a KMS state of the field, $\hat\rho_\beta$. For concreteness, we keep the generality in the spacetime localization of the operation but we choose \mbox{$\hat O(t, \bm{x}) = \lambda \chi(t) F(\bm{x}) \fieldtx$}. 
$\lambda$ regulates the strength of the process, $ \chi(t)\in\mathbb{R}$ is a time-switching supported over $t\in[\initialTime,\finalTime]$ and $F(\bm{x})\in\mathbb{R}$ localizes the process in space,
\begin{equation}
    \unitary = \mathcal{T} e^{-\ii \lambda \int_{\mathbb{R}}\!\differential t \chi(t) \int_{\mathbb{R}^n}\!\differential^n \bm{x}\, F(\bm{x}) \fieldtx }\,.
    \label{eq:UnitaryFamilyTimeordered}
\end{equation}
Performing a Magnus expansion and using the fact that \mbox{$[\field(t,\bm{x}), \field(t',\bm{x'})] \propto\openone$}, we simplify \eqref{eq:UnitaryFamilyTimeordered} to 
\begin{equation}
    \unitary = e^{\ii\theta} e^{-\ii\lambda\int_{\mathbb{R}}\!\differential t \chi(t) \int_{\mathbb{R}^n}\!\differential^n \bm{x}\, F(\bm{x}) \fieldtx } ,
    \label{eq:unitary_simplified}
\end{equation}
where the phase $e^{\ii\theta}$ does not affect time-evolution. In Appendix~\ref{apx:QuantumFieldsKMSCalculation} we compute the work and internal-energy-difference characteristic functions using Wick's theorem, yielding 
\begin{align}
    \widetilde{P}_\textsc{w}(\dvar) = \exp \left[\lambda^2  \int_{\mathbb{R}^n}\!\frac{\differential^n \bm{k} }{2(2\pi)^n\omega_{\bm{k}}} \abs{\widetilde{\chi}(\omega_{\bm{k}})}^2\abs*{\widetilde{F}(\bm{k})}^2 \right. \nonumber \\
    \left. \qty(\ii \sin{\omega_{\bm{k}}\mu} + \frac{e^{\beta \omega_{\bm{k}}}+1}{e^{\beta \omega_{\bm{k}}}-1}(\cos{\omega_{\bm{k}}\mu}-1))\right], \label{eq:characteristic_work_KMS}\\
    \widetilde{P}_{\scaleto{\operatorOfWork}{6pt}}(\mu) = \exp \left[\lambda^2  \int_{\mathbb{R}^n}\!\frac{\differential^n \bm{k} }{2(2\pi)^n} \abs{\widetilde{\chi}(\omega_{\bm{k}})}^2\abs*{\widetilde{F}(\bm{k})}^2 \right. \nonumber\\ 
    \left. \qty(\ii \mu - \frac{e^{\beta \omega_{\bm{k}}}+1}{e^{\beta \omega_{\bm{k}}}-1} \omega_{\bm{k}}\mu^2)\right], \label{eq:characteristic_DU_KMS}
\end{align}
where the tilde on a function notates its Fourier transform 
$\widetilde{F}(\bm{k})\!\coloneqq \int_{\mathbb{R}^n}\!\differential^n\bm{x}F(\bm{x}) e^{\ii \bm{k}\cdot\bm{x}}$.
From Eqs.~\eqref{eq:characteristic_work_KMS}~and~\eqref{eq:characteristic_DU_KMS}, we check that indeed the first law is satisfied in mean and variance
\begin{align}
 &\!\!\!\expval{W}_{\KMSstate}=\langle\operatorOfWork\rangle_{\KMSstate} = \lambda^2  \int_{\mathbb{R}^n}\!\frac{\differential^n \bm{k} }{2(2\pi)^n} \abs{\widetilde{\chi}(\omega_{\bm{k}})}^2\abs*{\widetilde{F}(\bm{k})}^2, \label{eq:workDUaveragevariance}\\
   \nonumber   & \sigma^2_{\textsc{w}}=\sigma^2_{\scaleto{\operatorOfWork}{5pt}} = \lambda^2  \int_{\mathbb{R}^n}\!\frac{\differential^n \bm{k} }{2(2\pi)^n} \abs{\widetilde{\chi}(\omega_{\bm{k}})}^2\abs*{\widetilde{F}(\bm{k})}^2  \omega_{\bm{k}} \frac{e^{\beta \omega_{\bm{k}}}+1}{e^{\beta \omega_{\bm{k}}}-1}. 
\end{align}
Notice that this is a non-perturbative expression (no small $\lambda$ assumption was made). 

The first law is not satisfied for higher moments. As shown in Appendix~\ref{apx:QuantumFieldsKMSCalculation} the exact expression for $\operatorOfWork$ is
\begin{align}
    \operatorOfWork = - \lambda \int_{\mathbb{R}}\!\differential t \chi(t) \int_{\mathbb{R}^n}\!\differential^n \bm{x}\, F(\bm{x}) \partial_t \fieldtx  \nonumber \\ 
    + \lambda^2 \int_{\mathbb{R}^n}\!\frac{\differential^n \bm{k} }{2(2\pi)^n}\abs{\widetilde{\chi}(\omega_{\bm{k}})}^2\abs*{\widetilde{F}(\bm{k})}^2\openone.
\end{align}
Hence, $\langle \operatorOfWork^j\rangle_{\state} \in \order{\lambda^j}$. On the other hand $\langle W^j\rangle_{\state_{_\beta}}$ always has terms proportional to $\lambda^2$, which makes the coincidence of higher moments of work and internal energy impossible. Notice that this is commensurate with the fact that the TPM distribution in finite dimensional systems cannot satisfy a first law in third or higher moments for all Gibbs states with either of our definitions of internal energy. Furthermore, \mbox{$ \langle W^j\rangle_{\state_{_\beta}} \geq \langle \operatorOfWork^j\rangle_{\state_{_\beta}}$} $\forall j\geq 3$.  In particular for the third moments 
\begin{equation}
\expval{W^3}_{\state_{_\beta}} - \big\langle \operatorOfWork^3\big\rangle_{\state_{_\beta}} = \lambda^2 \int_{\mathbb{R}^n}\!\frac{\differential^n \bm{k} }{2(2\pi)^n} \abs{\widetilde{\chi}(\omega_{\bm{k}})}^2\abs*{\widetilde{F}(\bm{k})}^2 \omega_{\bm{k}}^2 .
\end{equation}
This deviation grows when the process is more localized in space and time, (faster than the also growing $\langle \operatorOfWork^3\rangle$). For illustration, in Appendix~\ref{apx:QuantumFieldsKMSCalculation} we also show explicitly how Crooks theorem is satisfied in the present particular case.

\section{\textbf{Conclusion}} Standard definitions of work distributions in quantum thermodynamics do not work on quantum field theory. This is because of the lack of a notion of Gibbs thermality as well as the incompatibility of projective measurements with the relativistic nature of the QFT~\cite{Redhead1995, sorkin, Dowker, Dowker2,borsten2019impossible,Ortega2019work}. Taking this into account, we have extended the scope of quantum thermodynamics analyses to quantum field theory by 1) proposing a first law of thermodynamics for QFT that works even in the absence of Gibbs thermality and projective measurements, 2) identifying QFT-compatible work distributions that satisfy fluctuation theorems and the first law, and 3) presenting a full non-perturbative thermodynamic analysis of spacetime localized unitary processes on a quantum field. This analysis overcomes the limitations that emerge from applying finite-dimensional quantum thermodynamics to quantum field theory and paves the way to the use of the full power of quantum thermodynamics in QFT.

\acknowledgments
The authors are grateful to Nicole Yunger Halpern for very helpful discussions and her invaluable feedback. We also thank Jos\'e de Ram\'on for his very helpful insights. E. M-M. is supported by his Ontario Early Researcher Award and the NSERC Discovery program. A. T-B.  acknowledges the support of Fundaci\'o Privada Cellex, through a Mobility Research Award.

\appendix


\section{Comparison between definitions of the variance of internal energy difference}
\label{apx:InternalEnergyDifferenceDistribution}
The variances of the two internal energy difference distributions defined from  \eqref{eq:DUparallelMoments} and \eqref{eq:DUoperator} are related by
\begin{align}
\label{eq:secondmomentcomparison}
    \sigma^2_{\scaleto{\Delta\mathcal{U}}{3.7pt}} &= \sigma^2_{\scaleto{\operatorOfWork}{5pt}} + \expvalbig{\operatorOfWork\HO} +\expvalbig{\HO\operatorOfWork}\\\nonumber & -2\expvalbig{\operatorOfWork}\expvalbig{\HO} + 2 \expvalbig{\HO^2} -2\expvalbig{\HO}^2.  
\end{align}
Here we prove that the variance of $\operatorOfWork$ defined in \eqref{eq:DUoperator} is (for KMS states of a free scalar quantum field) the same as the variance in \eqref{eq:DUparallelMoments} after subtracting the divergent parts. Consider a KMS state $\KMSstate$ under the unitary process \eqref{eq:UnitaryFamilyTimeordered}. In this case \eqref{eq:secondmomentcomparison} becomes
\begin{align}
   \sigma^2_{\scaleto{\Delta\mathcal{U}}{3.7pt}} - \sigma^2_{\scaleto{\operatorOfWork}{5pt}}  &= 2 \big\langle{\HO^2}\big\rangle_{\state_{_\beta}} -2\big\langle{\HO}\big\rangle_{\state_{_\beta}}^2,
\end{align}
where we used $[\state_{_\beta},\HO]=0$, Eq.~\eqref{eq:DUoperatorExplicit} and that (since the KMS state is an even Gaussian state) the expectation of terms with different number of creation and annihilation operators vanishes. To compute the variance of \mbox{$\HO = \int_{\mathbb{R}^n}\!\differential^n \bm{k} \omega_{\bm{k}}\, \creation_{\bm{k}} \annihilation_{\bm{k}}$} for KMS states we use the canonical commutation relations  \mbox{$[\annihilation_{\bm{k}},\creation_{\bm{k}'}] = \delta(\bm{k} - \bm{k}')\openone$}, $\comm{\annihilation_{\bm{k}}}{\annihilation_{\bm{k}'}} = [\creation_{\bm{k}},\creation_{\bm{k}'}] = 0$, and the distributional expressions for the expectations of ladder operator products on KMS states (see e.g, in~\cite{Simidzija2018harvesting})
\begin{align}
  &  \Tr(\state_{_\beta}\creation_{\bm{k}}\annihilation_{\bm{k}'}) = \frac{1}{e^{\beta\omega_{\bm{k}}}-1} \delta(\bm{k} - \bm{k}') \label{eq:expectationCA},\\
  & \nonumber \Tr(\state_{_\beta}\Ck{1}\Ak{2}\Ck{3}\Ak{4}) = \frac{\delta(\k{3}-\k{4}) \delta(\k{1}-\k{2})}{(e^{\beta\omega_{\k{1}}}-1)(e^{\beta\omega_{\k{3}}}-1)}  \label{eq:expectationCACA} \\ &\qquad\qquad\quad+ \frac{e^{\beta\omega_{\k{3}}}\delta(\k{2}-\k{3}) \delta(\k{1}-\k{4})}{(e^{\beta\omega_{\k{1}}}-1)(e^{\beta\omega_{\k{3}}}-1)}.
\end{align}
Using \eqref{eq:expectationCA} and \eqref{eq:expectationCACA} we obtain
\begin{align}
     \sigma^2_{\scaleto{\Delta\mathcal{U}}{3.7pt}} - \sigma^2_{\scaleto{\operatorOfWork}{5pt}} &= 2\delta(0) \int\dk\frac{e^{\beta\omega_{\bm{k}}}}{(e^{\beta\omega_{\bm{k}}}-1)^2} \omega_{\bm{k}}^2. \label{eq:divergenceDU}
\end{align}
Which tells us that $\sigma^2_{\scaleto{\operatorOfWork}{5pt}}$ (which is finite as seen in \eqref{eq:workDUaveragevariance}) is equal to $\sigma^2_{\scaleto{\Delta\mathcal{U}}{3.7pt}}$ except for the divergent $\delta(0)$ term. In other words, $\sigma^2_{\scaleto{\operatorOfWork}{5pt}}$ can be thought of as the renormalized variance of \eqref{eq:DUparallelMoments}. Moreover, $\sigma^2_{\scaleto{\operatorOfWork}{5pt}}$ is a function of only the initial and final state, because \mbox{$\sigma^2_{\scaleto{\Delta\mathcal{U}}{3.7pt}} = f(\state,\U\state\U^\dagger)$} by its definition and $\sigma^2_{\scaleto{\Delta\mathcal{U}}{3.7pt}} - \sigma^2_{\scaleto{\operatorOfWork}{5pt}}$ does not depend on the process. In fact, the whole distribution of $\operatorOfWork$ is a function of the initial and the final state: the distribution is Gaussian (see Eq.~\eqref{eq:characteristic_DU_KMS}) and we already showed that the mean and variance are path independent.

Finally, as a quick note, we show that the TPM work distribution depends on the path taken from the initial to final state. We show this by giving the following example: choose $\HO=\HT = \epsilon\ketbra{1}{1}$,
$\state = \frac{1}{2}(\ketbra{0}{0}+\ketbra{1}{1})$, and two paths $\U = \openone$ and $\U'=\ketbra{0}{1}+\ketbra{1}{0}$. We have $\U\state\U^\dagger = \U'\state\U'^\dagger$, but the TPM distribution is different for each path: $\delta(W)$ and $\frac{1}{2}[\delta(W-\epsilon)+\delta(W+\epsilon)]$, respectively.

\begin{widetext}
\section{Comparison of higher moments of Work and Internal energy difference}
\label{apx:Moments_of_Work_and_EnergyDifference_of_KMS_Table}
For completion we provide the expressions for the mean and the central moments (up to the fourth) both for the distributions of $\operatorOfWork$ and $W$ for unitary processes of the form \eqref{eq:UnitaryFamilyTimeordered} applied on arbitrary KMS states of the field. The $j$-th central moment of $X$ is $\big\langle (X - \langle X \rangle)^j \big \rangle$. Notice that the distribution of internal energy difference is Gaussian and therefore all the odd central moments are zero.

\begin{table}[ht]
\centering
\bgroup
\def\arraystretch{2}
\begin{tabular}{|c|c|c|}
\hline
Moment & Internal energy difference                                                                                                    & Work (RS, FCS, ATMH) \\ \hline
Mean  $\langle X \rangle$ & $\lambda^2 \int_{\mathbb{R}^n}\!\frac{\differential^n \bm{k} }{2(2\pi)^n} \abs{\widetilde{\chi}(\omega_{\bm{k}})}^2\abs*{\widetilde{F}(\bm{k})}^2$  &   $\lambda^2 \int_{\mathbb{R}^n}\!\frac{\differential^n \bm{k} }{2(2\pi)^n} \abs{\widetilde{\chi}(\omega_{\bm{k}})}^2\abs*{\widetilde{F}(\bm{k})}^2$\\[1mm]
\hline
\makecell{Second central \\ $\big\langle (X - \langle X \rangle)^2 \big \rangle$} & $\lambda^2 \int_{\mathbb{R}^n}\!\frac{\differential^n \bm{k} }{2(2\pi)^n} \abs{\widetilde{\chi}(\omega_{\bm{k}})}^2\abs*{\widetilde{F}(\bm{k})}^2 \frac{e^{\beta \omega_{\bm{k}}}+1}{e^{\beta \omega_{\bm{k}}}-1} \omega_{\bm{k}}$             & $\lambda^2 \int_{\mathbb{R}^n}\!\frac{\differential^n \bm{k} }{2(2\pi)^n} \abs{\widetilde{\chi}(\omega_{\bm{k}})}^2\abs*{\widetilde{F}(\bm{k})}^2 \frac{e^{\beta \omega_{\bm{k}}}+1}{e^{\beta \omega_{\bm{k}}}-1} \omega_{\bm{k}}$ \\[2mm]
\hline
\makecell{Third central \\ $\big\langle (X - \langle X \rangle)^3 \big \rangle$}
&
0
& $\lambda^2 \int_{\mathbb{R}^n}\!\frac{\differential^n \bm{k} }{2(2\pi)^n} \abs{\widetilde{\chi}(\omega_{\bm{k}})}^2\abs*{\widetilde{F}(\bm{k})}^2 \omega_{\bm{k}}^2$ \\[2mm]

\hline
\makecell{Fourth central \\ $\big\langle (X - \langle X \rangle)^4 \big \rangle$}
&
    $3\lambda^4 \qty(\int_{\mathbb{R}^n}\!\frac{\differential^n \bm{k} }{2(2\pi)^n} \abs{\widetilde{\chi}(\omega_{\bm{k}})}^2\abs*{\widetilde{F}(\bm{k})}^2 \frac{e^{\beta \omega_{\bm{k}}}+1}{e^{\beta \omega_{\bm{k}}}-1} \omega_{\bm{k}})^2 $  
    &
    $\begin{aligned}
    &\lambda^2 \int_{\mathbb{R}^n}\!\frac{\differential^n \bm{k} }{2(2\pi)^n} \abs{\widetilde{\chi}(\omega_{\bm{k}})}^2\abs*{\widetilde{F}(\bm{k})}^2 \frac{e^{\beta \omega_{\bm{k}}}+1}{e^{\beta \omega_{\bm{k}}}-1} \omega_{\bm{k}}^3 \\
    &+3\lambda^4  \qty(\int_{\mathbb{R}^n}\!\frac{\differential^n \bm{k} }{2(2\pi)^n} \abs{\widetilde{\chi}(\omega_{\bm{k}})}^2\abs*{\widetilde{F}(\bm{k})}^2 \frac{e^{\beta \omega_{\bm{k}}}+1}{e^{\beta \omega_{\bm{k}}}-1} \omega_{\bm{k}})^2
    \end{aligned} $
     \\
\hline
        
\end{tabular}
\egroup
\end{table}

Finally let us show that this difference in higher moments is not unexpected. This is so because unlike the work distributions, neither of the internal energy difference distributions used in this paper fulfill the fluctuation theorems. This means that work and internal energy distributions have to differ for high enough moments. This fact is already known for the distribution associated to the operator $\Delta \hat{\mathcal{U}}$ \cite{Baumer2018proposals}. For the definition of  $\Delta \mathcal{U}$ that we give in Section~\ref{sec:Energy} we show that this is also the case by giving a counterexample to Jarzynski equality $\langle e^{-\beta\Delta\mathcal{U}}\rangle_{\KMSstate} = e^{-\beta \Delta F}$, which in turn, implies that Crooks theorem is not satisfied~\cite{Jarzynski1997nonequilibrium, sai2016thermodynamics}. Consider a two-dimensional quantum system with $\HO=\HT = \epsilon\ketbra{1}{1}$, an initial Gibbs state
$\state_{\beta} = \frac{1}{1+e^{-\beta\epsilon}}(\ketbra{0}{0}+e^{-\beta\epsilon}\ketbra{1}{1})$ and $\hat U = \openone$. Then,
\begin{equation}
    P(\Delta\mathcal{U})= \frac{(1+e^{-2\beta\epsilon})\delta(\Delta\mathcal{U}) + e^{-\beta\epsilon} [\delta(\Delta\mathcal{U}-\epsilon)+\delta(\Delta\mathcal{U}+\epsilon)]}{(1+e^{-\beta\epsilon})^2},
\end{equation}
from here $\langle e^{-\beta\Delta\mathcal{U}}\rangle_{\KMSstate} = 2\frac{1+e^{-2\beta\epsilon}}{(1+e^{-\beta\epsilon})^2}$. Together with \mbox{$\Delta F = 0$} due to $\HO=\HT$, this means Jarzynski equality does not hold $\langle e^{-\beta\Delta\mathcal{U}}\rangle_{\KMSstate}\neq 1$, and therefore neither does Crooks theorem.

\end{widetext}

\section{Details of the work and internal energy difference statistics of processes on quantum fields}
\label{apx:QuantumFieldsKMSCalculation}
\subsection{Closed-form expression for the family of unitary processes}
\label{apx:QF_ClosedFormFamilyLocalizedUnitaries}
We will need to make use of the mode expansion of the field operator:
\begin{align}
    \field(t,\bm{x}) = \int_{\mathbb{R}^n}\!\frac{ \differential^n \bm{k}}{\sqrt{2(2\pi)^n \omega_{\bm{k}}}} \qty(e^{-\ii \mathsf{k} \cdot \mathsf{x}} \creation_{\bm{k}} + e^{\ii \mathsf{k} \cdot \mathsf{x}} \annihilation_{\bm{k}}) ,
\\
    \mathsf{k} \cdot \mathsf{x} = \bm{k} \cdot \bm{x} - \omega_{\bm{k}}t \,, \quad \omega_{\bm{k}} = \sqrt{m^2 + \vert\bm{k}\vert^2}\,.
\end{align}
Where  we denote with $m$ the mass of the field.
For convenience, we define:
\begin{equation}
    \field_F(t) \coloneqq \int_{\mathbb{R}^n}\!\differential^n \bm{x}\, F(\bm{x}) \fieldtx, \quad \field_{\chi F} \coloneqq \int_{\mathbb{R}}\!\differential t\, \chi(t) \field_F(t).
\end{equation}
The unitary processes that we apply to the field, as stated in \eqref{eq:UnitaryFamilyTimeordered}, is
\begin{equation}
    \unitary = \mathcal{T} e^{-\ii \lambda \field_{\chi F}},
    \label{eq:timeOrderedUnitaryAppendix}
\end{equation}
Commonly in QFT we would take a Dyson expansion and perform an infinite series of nested time integrals. However, we can fully sum the series by performing a Magnus expansion instead. This is possible because $\commsmall{\field_F(t)}{\field_F(t')} \propto \openone$, which causes further commutators with $\field_F$ to vanish. The consequence is that we obtain the following closed-form for the family of unitaries
\begin{equation}
    \unitary = e^{\ii\theta} e^{-\ii\lambda\field_{\chi F}} ,
    \label{eq:closed-formUnitaries}
\end{equation}
where 
\begin{align}
    &\theta = \ii \lambda^2\int_{\mathbb{R}}\!\differential t \int_{-\infty}^t\!\!\!\! \differential t'\chi(t)\chi(t') \expvalstate{\comm{\field_F(t)}{\field_F(t')}} \\
    &\!\!= \lambda^2\int_{\mathbb{R}^n}\!\frac{ \differential^n \bm{k} \abs*{\widetilde{F}(\bm{k})}^2}{(2\pi)^n \omega_{\bm{k}}} \int_{\mathbb{R}}\!\differential t \int_{-\infty}^t\!\!\!\! \differential t'\chi(t)\chi(t') \sin\left[\omega_{\bm{k}} (t-t')\right]. \nonumber
\end{align}
$\hat U$ is a displacement operator (up to an irrelevant phase) \mbox{$\displacement_{\alpha(\bm k)}=\exp[\int_{\mathbb{R}^n}\text{d}^n\bm k\, (\alpha(\bm k)\hat a^\dagger -\alpha^*(\bm k)\hat a)]$}. The displacement operator coherent amplitude distribution is frequency dependent similar to the one found in~\cite{Simidzija2018harvesting}. Concretely,
\begin{equation}
    \alpha(\bm{k}) = -\ii\lambda \frac{\widetilde{\chi}(\omega_{\bm{k}})\widetilde{F}(\bm{k})^*}{\sqrt{2(2\pi)^n \omega_{\bm{k}}}}.
\end{equation} 
The action of this displacement operator on an annihilation operator of a mode of wavevector $\bm k$ is
\begin{equation}
    \hat U_{\alpha}^\dagger \annihilation_{\bm{k}} \hat U_\alpha = \annihilation_{\bm{k}} + \alpha(\bm{k})\openone\,.
    \label{eq:evolution_annihilation}
\end{equation}

\subsection{Internal energy difference distribution for KMS states}
\label{apx:QF_EnergyDifferenceKMSCalculation}
Here we calculate the characteristic function of internal energy difference for initial KMS states, which will give us access to its moments. 

First, we compute a general non-perturbative expression for \mbox{$\operatorOfWork = \unitaryDagger\HO\unitary - \HO$}. This is for a process that starts and ends with the free field Hamiltonian, \mbox{$\HO = \int_{\mathbb{R}^n}\!\differential^n \bm{k} \omega_{\bm{k}}\, \creation_{\bm{k}} \annihilation_{\bm{k}}$}. Using Eq.\,\eqref{eq:evolution_annihilation} we get
\begin{align}
    \unitaryDagger\HO\unitary =\int_{\mathbb{R}^n}\!\differential^n \bm{k} \omega_{\bm{k}} \Big(&\creation_{\bm{k}}\annihilation_{\bm{k}} + \alpha(\bm{k})\creation_{\bm{k}}   \\  &+ \alpha(\bm{k})^*\annihilation_{\bm{k}} + \abs{\alpha(\bm{k})}^2\openone\Big).\nonumber
\end{align}
We identify that $\int_{\mathbb{R}^n}\!\differential^n \bm{k} \omega_{\bm{k}}(\alpha(\bm{k})\creation_{\bm{k}} + \alpha(\bm{k})^*\annihilation_{\bm{k}})$ is proportional to 
\begin{align}
\hat\pi_{\chi F}&\coloneqq\int_\mathbb{R}\!\!\differential t\,\chi(t) \partial_t\field_{F}(t) \\ \nonumber
&=\int_{\mathbb{R}^n}\!\frac{ \differential^n \bm{k} \sqrt{\omega_{\bm{k}}}}{\sqrt{2(2\pi)^{n}}} \qty(\ii\widetilde{\chi}(\omega_{\bm{k}}) \widetilde{F}(\bm{k})^*\creation_{\bm{k}} -\ii \widetilde{\chi}(\omega_{\bm{k}})^* \widetilde{F}(\bm{k})\annihilation_{\bm{k}}).
\end{align}
Hence, the internal energy difference operator can be written as 
\begin{align}
    \operatorOfWork 
    &= - \lambda \hat\pi_{\chi F} + \lambda^2 \int_{\mathbb{R}^n}\!\frac{\differential^n \bm{k} }{2(2\pi)^n}\abs{\widetilde{\chi}(\omega_{\bm{k}})}^2\abs*{\widetilde{F}(\bm{k})}^2\openone\,. \label{eq:DUoperatorExplicit}
\end{align}

The characteristic function for the internal energy difference in a KMS state is~\cite{Talkner2007fluctuationtheorems}
\begin{equation}
    \widetilde{P}_{\operatorOfWork}(\dvar) = \expval{e^{\ii\dvar \operatorOfWork}}_{\state_{_\beta}} ,
    \label{eq:characteristic_general_DU}
\end{equation}
where recall that $\beta$ indicates the inverse temperature of the KMS state. The term in $\operatorOfWork$ proportional to the identity will give a state-independent contribution. We still need to evaluate $\langle{e^{- i \dvar \lambda \hat\pi_{\chi F}}}\rangle_{\state_{_\beta}}$. Using Wick's theorem and the fact that a KMS state is quasi-free (Gaussian with a zero one-point function) we obtain~\cite{ramn2020nonperturbative}
\begin{equation}
    \expval{e^{- i \dvar \lambda \hat\pi_{\chi F}}}_{\state_{_\beta}} \!\! = e^{- \frac{1}{2} \dvar^2 \lambda^2 \expval{\qty(\hat\pi_{\chi F})^2}_{\state_{_\beta}}}.
\end{equation}
The expectation $\langle{(\hat\pi_{\chi F})^2}\rangle_{\state_{_\beta}}$ is given by time derivatives of \mbox{$\mathcal{W}_{\state_{_\beta}}(t,x,t',\bm{x}') \coloneqq \langle{\fieldtx \field(t',\bm{x}')}\rangle_{\state_{_\beta}}$}, the KMS state Wightman function. Namely, 
\begin{align}
    \expval{\qty(\hat\pi_{\chi F})^2}_{\state_{_\beta}} = &  \int_{\mathbb{R}}\!\differential t \chi(t) \int_{\mathbb{R}^n}\!\differential^n \bm{x} F(\bm{x}) \int_{\mathbb{R}}\!\differential t' \chi(t')  \nonumber \\ & \times  \int_{\mathbb{R}^n}\!\differential^n \bm{x}' F(\bm{x}') \partial_{t} \partial_{t'} \mathcal{W}_\beta(t,x,t',\bm{x}').
\end{align}
The Wightman function for a KMS state of a free scalar field is well known (see, e.g.,~\cite{Simidzija2018harvesting,Strocchi2008}):
\begin{align}
    \mathcal{W}_\beta(t,x,t',\bm{x}') = &\int_{\mathbb{R}^n}\!\frac{ \differential^n \bm{k}}{2(2\pi)^n \omega_{\bm{k}}(e^{\beta \omega_{\bm{k}}}-1)} \nonumber \\&\times \qty(e^{\beta \omega_{\bm{k}}}e^{\ii\mathsf{k}\cdot(\mathsf{x}-\mathsf{x}')} + e^{-\ii\mathsf{k}\cdot(\mathsf{x}-\mathsf{x}')}).
    \label{eq:thermalWightman}
\end{align}
Gathering all the results above, the non-perturbative expression of the characteristic function of internal energy difference yields
\begin{align}
    \widetilde{P}_{\operatorOfWork}(\mu) = \exp \left[ \lambda^2  \int_{\mathbb{R}^n}\!\frac{\differential^n \bm{k} }{2(2\pi)^n} \abs{\widetilde{\chi}(\omega_{\bm{k}})}^2\abs*{\widetilde{F}(\bm{k})}^2 \right.\nonumber \\ \left. \times \qty(\ii \mu   - \frac{1}{2} \dvar^2 \omega_{\bm{k}} \frac{e^{\beta \omega_{\bm{k}}}+1}{e^{\beta \omega_{\bm{k}}}-1})\right].
    \label{eq:CharacteristicInternalEnergy_KMS_appendix}
\end{align}
\subsection{Work distribution for KMS states}
\label{apx:QF_WorkKMSCalculation}
Here we calculate the work characteristic function for KMS states. The expression for the characteristic function is given by Eq.~\eqref{eq:RamseyCharacteristicWork}. First we consider the general family of processes defined by  Eq.~\eqref{eq:UnitaryFamilyTimeordered} whose close expression is given in Eq.~\eqref{eq:closed-formUnitaries}. Using~\eqref{eq:closed-formUnitaries} we get
\begin{equation}
\widetilde{P}_{\textsc{w}}(\dvar) = \expval{e^{\ii\lambda\field_{\chi F}} e^{\ii\dvar\HO} e^{-\ii\lambda\field_{\chi F}} e^{-\ii\dvar\HO}}_{\state_{_\beta}} .
\label{eq:characteristic_work_KMS_as_expectation_apx}
\end{equation}
Same as we did before, since the KMS state is quasi-free (with zero one-point function) we can easily use Wick's theorem to evaluate the expectation value. It is convenient to first show that
\begin{equation}
\begin{split}
    &e^{\ii\dvar\HO} e^{-\ii\lambda\field_{\chi F}} e^{-\ii\dvar\HO}   \\ & = \exp(-\ii\lambda \int_{\mathbb{R}}\!\differential t \chi(t) e^{\ii\dvar\HO}\field_{F}(t)e^{-\ii\dvar\HO}) \\
    & = \exp(-\ii\lambda \int_{\mathbb{R}}\!\differential t \chi(t) \field_{F}(t+\mu)) \\
    & = e^{-\ii\lambda\field_{\gamma F}},
\end{split}
\end{equation}
with $\gamma(t) = \chi(t-\mu)$. We use the BCH formula to get:
\begin{equation}
    \expval{e^{\ii\lambda\field_{\chi F}} e^{-\ii\lambda\field_{\gamma F}}}_{\state_{_\beta}}\!\! = e^{\frac{1}{2}\lambda^2 \expvalstate{\comm{\field_{\chi F}}{\field_{\gamma F}}}}\expval{e^{\ii\lambda(\field_{\chi F}-\field_{\gamma F})}}_{\state_{_\beta}}\!\!,
\end{equation}
where again the higher order commutators vanish since $\commsmall{\field_F(t)}{\field_F(t')} \propto \openone$. Now we use that the state is Gaussian with zero one-point function to get~\cite{ramn2020nonperturbative}
\begin{equation}
    \expval{e^{\ii\lambda(\field_{\chi F}-\field_{\gamma F})}}_{\state_{_\beta}} = e^{-\frac{1}{2}\lambda^2\expval{\qty(\field_{\chi F}-\field_{\gamma F})^2}_{\state_{_\beta}}}.
\end{equation}
With this, we will get the characteristic function in terms of the Wightman function, as in the former calculation for the internal energy difference. $\langle{(\field_{\chi F})^2}\rangle_{\state_{_\beta}}$ and $\langle{(\field_{\gamma F})^2}\rangle_{\state_{_\beta}}$ have the same value, because the KMS state is stationary and $\gamma$ is a time-shifted version of $\chi$. Putting everything together, the characteristic function is
\begin{equation}
    \expval{e^{\ii\lambda\field_{\chi F}} e^{-\ii\lambda\field_{\gamma F}}}_{\state_{_\beta}} = e^{\lambda^2\qty(\expval{\field_{\chi F}\field_{\gamma F}}_{\state_{_\beta}}-\expval{(\field_{\chi F})^2}_{\state_{_\beta}})}\,,
\end{equation}
where in terms of Wightman functions,
\begin{align}
   &\expval{\field_{\chi F}\field_{\gamma F}}_{\state_{_\beta}}-\expval{(\field_{\chi F})^2}_{\state_{_\beta}}  \\ &=\int_{\mathbb{R}}\!\differential t \chi(t) \int_{\mathbb{R}^n}\!\differential^n \bm{x} F(\bm{x}) \int_{\mathbb{R}}\!\differential t' \chi(t') \int_{\mathbb{R}^n}\!\differential^n \bm{x}' F(\bm{x}') \nonumber \\ 
   &\quad\times (\mathcal{W}_\beta(t,x,t'+\mu,\bm{x}') - \mathcal{W}_\beta(t,x,t',\bm{x}')). \nonumber
\end{align}
The exact expression for the characteristic function of work for thermal states arises after substituting the value of the KMS Wightman functions given in \eqref{eq:thermalWightman},
\begin{align}
    \widetilde{P}_{\textsc{w}}(\dvar) = \exp\left[\lambda^2  \int_{\mathbb{R}^n}\!\frac{\differential^n \bm{k} }{2(2\pi)^n\omega_{\bm{k}}} \abs{\widetilde{\chi}(\omega_{\bm{k}})}^2\abs*{\widetilde{F}(\bm{k})}^2 \right. \nonumber \\ \left. \times \qty(\ii \sin{\omega_{\bm{k}}\mu} + \frac{e^{\beta \omega_{\bm{k}}}+1}{e^{\beta \omega_{\bm{k}}}-1}(\cos{\omega_{\bm{k}}\mu}-1))\right].
    \label{eq:CharacteristicWork_KMS_appendix}
\end{align}
It is easy to check that for small $\lambda$, this non-perturbative expression matches with the leading order perturbative expression found in~\cite{Ortega2019work}. 

\subsection{Proof that the moments of work are bigger than the moments of internal energy difference}
\label{apx:positive_moments_work_bigger_energy}
We now overview the proof that \mbox{$\langle W^j\rangle_{\state_{_\beta}} \geq \langle \operatorOfWork^j\rangle_{\state_{_\beta}}$} \mbox{$\forall j \geq 1$}, for unitaries of the form \eqref{eq:UnitaryFamilyTimeordered}. As a side result we get $\langle W^j\rangle_{\state_{_\beta}},\langle \operatorOfWork^j\rangle_{\state_{_\beta}} \geq 0\ \forall j \geq 1$. We use
\begin{equation}
\langle X^j\rangle_{\state_{_\beta}} = \ii^{-j} \eval{\frac{\differential^j}{\differential\dvar^j}\widetilde{P}_X(\dvar)}_{\dvar=0} ,\quad X \in\{ W,\, \operatorOfWork\},
\end{equation}
with the $\widetilde{P}_\textsc{w}$ and $\widetilde{P}_{\operatorOfWork}$ of \eqref{eq:CharacteristicWork_KMS_appendix} and \eqref{eq:CharacteristicInternalEnergy_KMS_appendix}.
First we show that $\ii^{-j} \frac{\differential^j}{\differential\dvar^j}\widetilde{P}_\textsc{x}(\dvar)$
can be expressed as a sum of positive terms of the form
\begin{align}
    c\widetilde{P}_{X}(\mu)\prod_{l=0}^j \left(\ii^{-l}\frac{\differential^l}{\differential\dvar^l}\alpha_{X}(\dvar)\right)^{i_l}, \quad X \in \{W,\, \operatorOfWork\}. \label{eq:summands_characteristic_derivatives}
\end{align}
Where $e^{\alpha_{X}(\mu)} = \widetilde{P}_{X}(\mu)$, $i_l,c \in \mathbb{Z}$, $i_l \geq 0,\, c \geq 1$. The proof goes by induction over $j$. The base case, $j=0$, is trivial. For $j>0$ we assume that $\ii^{-j}\frac{\differential^j}{\differential\dvar^j}\widetilde{P}(\dvar)$ is a sum of terms of the form of \eqref{eq:summands_characteristic_derivatives} and show the correspondent statement for $j+1$. We can check that it is true applying $\ii^{-1}\frac{\differential}{\differential\dvar}$ to \eqref{eq:summands_characteristic_derivatives} which returns a sum of terms with the same form of \eqref{eq:summands_characteristic_derivatives} but with $j$ increased by one.

The $j$-th moments of $W$ and $\operatorOfWork$ are the sum of terms of the form of \eqref{eq:summands_characteristic_derivatives} at $\mu = 0$, which are all positive because
\begin{align}
    \widetilde{P}_X(0) = 1,
    \ 
    \eval{\ii^{-l}\frac{\differential^l}{\differential\dvar^l}\alpha_X(\dvar)}_{\dvar=0}\!\!\!\!\!\!\! \geq 0,\; \forall l \geq 0,\, X \in\{ W,\, \operatorOfWork\}.
\end{align}
We finish the proof showing that each summand of $\langle W^j\rangle_{\state_{_\beta}}$ is equal or  larger than the corresponding one in $\langle \operatorOfWork^j\rangle_{\state_{_\beta}}$, which is implied by
\begin{align}
    \eval{\ii^{-l}\frac{\differential^l}{\differential\dvar^l}\alpha_{\textsc{w}}(\dvar)}_{\dvar=0} \geq \eval{\ii^{-l}\frac{\differential^l}{\differential\dvar^l}\alpha_{\operatorOfWork}(\dvar)}_{\dvar=0} \quad l \geq 1.
\end{align}
Consequently, $\langle W^j\rangle_{\state_{_\beta}} \geq \langle \operatorOfWork^j\rangle_{\state_{_\beta}} \ \forall j \geq 1$.

\vspace{-2mm}

\subsection{Particular case of Crooks theorem}
\label{apx:QFT_KMS_Crooks_example}
Although we showed that Crooks theorem is satisfied in general, we show here for illustration that the $\widetilde{P}_{\textsc{w}}$ in \eqref{eq:CharacteristicWork_KMS_appendix} explicitly fulfills Crooks theorem \eqref{eq:CrooksTheorem}. Crooks theorem takes the form $\widetilde{P}_\textsc{w}(\dvar + \ii \beta) = \widetilde{P}_{\text{rev}}(-\dvar)$, because  \mbox{$\chi(t\notin[0,\tau])=0\Rightarrow \HO = \HT$} and therefore $\Delta F = 0$. We need to evaluate $\widetilde{P}_{\text{rev}}$.  
We see $ \widetilde{P}_{\text{rev}}=\widetilde{P}_\textsc{w}$ in our case by looking at \eqref{eq:characteristic_work_KMS_as_expectation_apx} and realizing that $\widetilde{P}_{\text{rev}}$ is the same as $\widetilde{P}_{\textsc{w}}$ but exchanging the variable $\lambda$ for $-\lambda$, which does not  change its expression \eqref{eq:CharacteristicWork_KMS_appendix}.
Furthermore, we convert the sine and cosine of \eqref{eq:CharacteristicWork_KMS_appendix} to sums of imaginary exponentials to obtain
\begin{align}
    \widetilde{P}_\textsc{w}(\dvar) = \exp \left[\lambda^2  \int_{\mathbb{R}^n}\!\frac{\differential^n \bm{k} }{2(2\pi)^n\omega_{\bm{k}}} \abs{\widetilde{\chi}(\omega_{\bm{k}})}^2\abs*{\widetilde{F}(\bm{k})}^2 \right. \nonumber \\ \left. \times \frac{(1-e^{-\ii \omega_{\bm{k}}\mu}) \qty(e^{\ii \omega_{\bm{k}}(\mu -\ii\beta) }-1)}{e^{\beta \omega_{\bm{k}}}-1}\right],
\end{align}
which clearly fulfills $\widetilde{P}_\textsc{w}(\dvar + \ii \beta) = \widetilde{P}_\textsc{w}(-\dvar)$ and consequently Crooks theorem.

\bibliography{bibliography}

\end{document}